\documentclass[conference, 10pt]{IEEEtran}
\usepackage{adjustbox}
\usepackage{amsmath}
\usepackage{cleveref}
\usepackage{comment}
\usepackage{enumitem}
\usepackage{graphicx}
\usepackage{multirow}
\usepackage{threeparttable}
\usepackage{subfig}
\usepackage{soul}
\usepackage{float}
\usepackage{booktabs}
\usepackage{xargs}
\usepackage{array}
\usepackage{bm}
\usepackage{verbatim}
\usepackage{caption}
\usepackage{subcaption}
\usepackage{graphicx}
\usepackage{url}
\usepackage{multicol,tabularx,capt-of}
\usepackage[svgnames]{xcolor}
\newcolumntype{L}[1]{>{\raggedright\arraybackslash}p{#1}}
\newcolumntype{C}[1]{>{\centering\arraybackslash}p{#1}}
\newcolumntype{R}[1]{>{\raggedleft\arraybackslash}p{#1}}
\newcolumntype{J}[1]{>{\justifying\arraybackslash}p{#1}}
\usepackage{amssymb}
\usepackage{pifont}
\newcommand{\cmark}{\ding{51}}%
\newcommand{\xmark}{\ding{56}}%
\def\BibTeX{{\rm B\kern-.05em{\sc i\kern-.025em b}\kern-.08em
    T\kern-.1667em\lower.7ex\hbox{E}\kern-.125emX}}
\usepackage{fancyhdr}
\usepackage{cite}

\AtBeginDocument{%
  \providecommand\BibTeX{{%
    \normalfont B\kern-.05em{\sc i\kern-.025em b}\kern-.08em\TeX}}}

\begin{document}

\title{IMPLY-based Approximate Full Adders for Efficient Arithmetic Operations in \\ Image Processing and Machine Learning}
\author{Melanie Qiu$^+$, Caoyueshan Fan$^+$, Gulafshan$^+$, Salar Shakibhamedan$^*$, Fabian Seiler$^*$ and Nima TaheriNejad$^{+*}$ \\
$^+$Heidelberg University, Germany, $^*$Technische Universität Wien (TU Wien), Austria \\
melanie.qiu@stud.uni-heidelberg.de, caoyueshan.fan@stud.uni-heidelberg.de, gulafshan@ziti.uni-heidelberg.de, \\ salar.shakibhamedan@tuwien.ac.at, fabian.seiler@tuwien.ac.at, nima.taherinejad@ziti.uni-heidelberg.de}

\maketitle

\begin{abstract}
To overcome the performance limitations in modern computing, such as the power wall, emerging computing paradigms are gaining increasing importance.
Approximate computing offers a promising solution by substantially enhancing energy efficiency and reducing latency, albeit with a trade-off in accuracy.
Another emerging method is memristor-based In-Memory Computing (IMC) which has the potential to overcome the Von Neumann bottleneck. 
In this work, we combine these two approaches and propose two \textbf{S}erial \textbf{APP}roximate \textbf{I}MPLY-based full adders (SAPPI). 
When embedded in a Ripple Carry Adder (RCA), our designs reduce the number of steps by 39\%-41\% and the energy consumption by 39\%-42\% compared to the exact algorithm.
We evaluated our approach at the circuit level and compared it with State-of-the-Art (SoA) approximations where our adders improved the speed by up to 10\% and the energy efficiency by up to 13\%.
We applied our designs in three common image processing applications where we achieved acceptable image quality with up to half of the RCA approximated.

We performed a case study to demonstrate the applicability of our approximations in Machine Learning (ML) underscoring the potential gains in more complex scenarios.
The proposed approach demonstrates energy savings of up to 296~mJ (21\%) and a reduction of 1.3 billion (20\%) computational steps when applied to Convolutional Neural Networks (CNNs) trained on the MNIST dataset while maintaining accuracy.
\end{abstract}

\section{Introduction}

Reducing energy consumption is one of the most relevant challenges in Information and Communication Technology (ICT). 
A large portion of the energy budget of modern computing devices (e.g. 63\% at Google~\cite{Boroumand2018GoogleEnergy}) is consumed by data movement alone, which totals to 1.07PWh of global energy~\cite{IMCTaheriNejad2024}. 
Data movement, constrained by the Von Neumann bottleneck, is a key limiting factor in throughput, making it an optimal target for performance enhancements.
With In-Memory Computing (IMC) the computation takes place directly in the memory, offering a potential solution to this issue. 
Memristors stand out as ideal components for IMC, since they can store data non-volatile with their resistive states and perform logical operations. Other advantages such as low power consumption, a small form factor, and compatibility with CMOS technology emphasize the memristor as a promising candidate~\cite{TaheriNejad2015, TaheriNejad2016, Radakovits2019, Borghetti2010MemSw, Lehtonen2009StatefulIL, Strukov2008TheMM}. Plenty of stateful logic forms based on memristors were proposed such as~\cite{Gupta2018Felix, TaheriNejad2021SIXOR, Huang2016tmsl, Kvatinsky2014MAGIC, Kvatinsky2011IMPLY}. Memristor-based Material Implication (IMPLY) proves to be a favorable choice, as it is well established, compatible with the crossbar array, and more reliable when compared to other stateful logic forms~\cite{Radakovits2021BELIEVER, Borghetti2010MemSw, Lehtonen2009StatefulIL}.
Approximate Computing (AxC) is another emerging computer paradigm that presents a possible solution to the power-wall problem~\cite{Liu2020AppComp, Gupta2013DSP, Jiang2017AxReview}. With AxC important metrics like energy efficiency and latency can be improved drastically. The trade-off for this is reduced accuracy. Since applications such as image and video processing, as well as Machine Learning (ML) are inherently error-resilient, they provide the optimal environment for AxC. As in modern computing, most of the operations are based on addition and multiplication, improving them with approximations is the most beneficial approach. 

In this work, we propose two IMPLY-based approximated adder algorithms for the serial topology (SAPPI-1\&2) that minimize the energy consumption and number of computational cycles by utilizing the unique properties of IMPLY. We present the first adder algorithm that maintains all input states after the operation and analyze the applicability in three image processing tasks and two machine learning tasks. 
This paper is structured as follows: In Section II we review the related literature.
Section III introduces the proposed approximate adders. In Section IV we simulate and analyze the adders and compare them to the SoA in Section V. The results of three image processing applications and their quality is discussed in Section VI. In Section VII, we show the applicability of our approach in machine learning, and in Section VIII, we conclude the paper.

\section{IMPLY-based Approximate Adders}

\begin{figure}[tb]
  \centering
  \begin{minipage}{0.48\columnwidth}
    \centering
    \includegraphics[width=0.48\columnwidth]{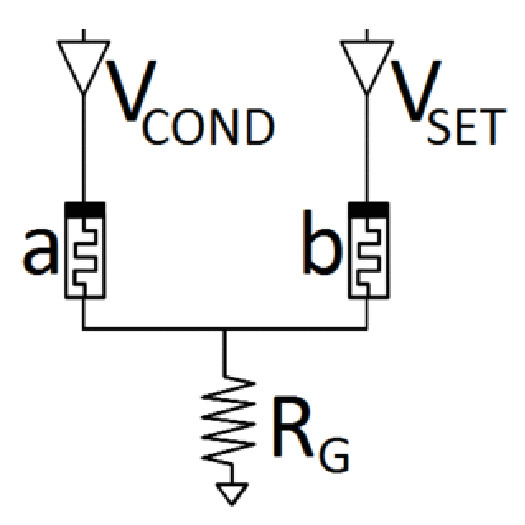}
  \end{minipage}
  \hfill
  \begin{minipage}{0.48\columnwidth}
    \centering
    \resizebox{0.8\columnwidth}{!}{
    \begin{tabular}{|c|c|c|}
        \hline
        \textbf{P} & \textbf{Q} & $\bm{P \text{ IMP } Q \equiv P \rightarrow Q}$ \\
        \hline \hline
        0 & 0 & 1  \\ 
        0 & 1 & 1  \\
        1 & 0 & 0  \\
        1 & 1 & 1  \\
        \hline
    \end{tabular}}
  \end{minipage}

  \vspace{1em}

  \begin{minipage}{\columnwidth}
    \centering
    \includegraphics[width=\columnwidth]{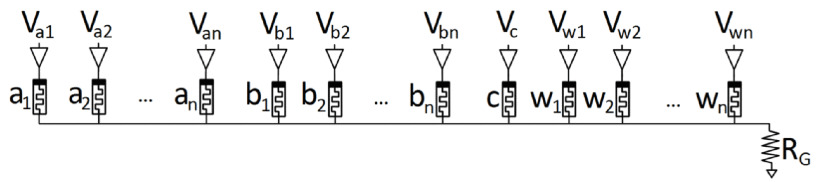}
  \end{minipage}

  \caption{IMPLY operation: (top) Basic Gate and Truth Table, (bottom) Serial Topology \cite{Fatemieh2023AFAIP, Borghetti2010MemSw}
}

  \label{fig:IMPLY}

\end{figure}

Memristors, originally discovered by L. Chua~\cite{Chua1971MM} and physically realized by R. Williams et al.~\cite{Strukov2008TheMM}, can store logical values via electrical resistance in a non-volatile fashion. They have low power consumption and small dimensions, deeming them the ideal memory cell~\cite{Borghetti2010MemSw, Lehtonen2009StatefulIL}. It is the convention to assume the minimum ($R_{on}$) and maximum ($R_{off}$) resistance as logical `1' and `0' which can be set by the applied voltage and the direction of the current~\cite{Chua1971MM, Strukov2008TheMM, Rohani2017MemFA, Karimi2018FullAdder, Radakovits2020MemristiveMultiplier}.
Stateful logic forms based on memristors include various models such as~\cite{Gupta2018Felix, TaheriNejad2021SIXOR, Huang2016tmsl, Kvatinsky2014MAGIC}. In this work, we focus on IMPLY~\cite{Kvatinsky2011IMPLY, Kvatinsky2014IMPLY} due to its reliability and the presence of other approximations~\cite{Radakovits2021BELIEVER, Fatemieh2023AFAIP, Seiler2023SSAxA, Fatemieh2022AppIMC}.
An IMPLY operation is denoted as $P \rightarrow Q$ and follows the truth table in~\Cref{fig:IMPLY}. The result is stored in the $Q$-memristor, overwriting the initial state.
The IMPLY structure (\Cref{fig:IMPLY}) consists of two memristors connected to a resistor where the resistances must satisfy the condition $R_{on} << R_{G} << R_{off}$. Voltages must meet $V_{COND} < V_{C} < V_{SET}$ for correct operation, where $V_{C}$ denotes the threshold voltage of the memristors~\cite{Borghetti2010MemSw, Kvatinsky2011IMPLY, Kvatinsky2014IMPLY, Karimi2018FullAdder}.
IMPLY-based full adders can be categorized into serial~\cite{Borghetti2010MemSw, Lehtonen2009StatefulIL, Rohani2017MemFA}, parallel~\cite{Karimi2018FullAdder, Kvatinsky2014IMPLY} and hybrid forms such as semi-serial~\cite{TaheriNejad2019SemiSerial,Radakovits2020MemristiveMultiplier} and semi-parallel~\cite{Rohani2020SemiParallel}. 
In this work we utilize the serial topology, illustrated in~\Cref{fig:IMPLY}, as it stands out for its low hardware complexity and small area, requiring only $2n+3$ memristors and no additional switches to function~\cite{Rohani2017MemFA}. 
Approximate computing allows for significant reductions in energy consumption and speed albeit at the cost of reduced accuracy. This technique is suitable for error-resilient applications like image processing and machine learning~\cite{Jiang2017AxReview, Jiang2020AAC, Almurib2016AppDes, Fatemieh2023AFAIP}. 
The resulting accuracy is evaluated using error metrics such as the Error Distance (ED), Mean Error Distance (MED), Normalized Mean Error Distance (NMED), and Mean Relative Error Distance (MRED)\cite{Jiang2017ARC, Jiang2017AxReview, Fatemieh2023AFAIP}. In image processing, quality metrics such as the Peak Signal-to-Noise Ratio (PSNR) and Mean Structural Similarity Index (MSSIM) are used to assess the resulting quality.
In the literature, a PSNR over 30dB is considered acceptable quality~\cite{Mittal2016ApproxComputing}.
IMPLY-based approximate adders have been proposed in both serial~\cite{Fatemieh2023AFAIP, Asgari2024SAFAN} and semi-serial topologies~\cite{Seiler2023SSAxA}, focusing on minimizing error metrics for image processing. Here, we propose two serial adders, SAPPI-1 and SAPPI-2, designed to optimize energy consumption and computational steps while maintaining acceptable quality for image processing
and accuracy in machine learning.

\section{Proposed Approximate Full Adders}
\label{Section3}

\subsection{Design Methodology}

The design of SAPPI-1 and SAPPI-2 is driven by the need to balance computational complexity and accuracy while ensuring energy efficiency and fast processing. The primary goal is to present fast and energy-efficient adders that deliver acceptable quality in applications like image processing and machine learning.
In the serial IMPLY topology, only one IMPLY or FALSE operation can be executed per cycle. Since these two functions form the complete logic set $\{\rightarrow, \perp\}$, all boolean operations can be emulated using them. Rather than directly implementing boolean expressions in IMPLY, which is inefficient~\cite{Rohani2017MemFA, Seiler2023SSAxA}, we utilize the unique properties of IMPLY logic.
In IMPLY logic, the result of the operation overwrites the initial state of an input, losing the stored information. With our methodology, we propose an approach to calculate the outputs without losing the initial states of the input memristors. When the inputs are required for multiple calculations our approach does not require an inefficient reloading process of the input data. We propose two approaches that utilize the work memristor $M$ to accumulate information from both inputs while preserving the input information. For this, we sequentially implicate the content of the $A$-memristor and the $B$-memristor to the initially reset (set to `0') work memristor, which represents the $Sum$ output. The $C_{out}$ is calculated by overwriting the $C$ input by implicating the $Sum$ output to $C$-memristor. With this, the carry-out is only erroneous in a single place, which reduces the incorrect error propagation.
We will denote both of our approaches as \textbf{S}erial \textbf{APP}roximate \textbf{I}MPLY-based full adders (SAPPI-1\&2).
In our first approach (SAPPI-1), both inputs are preserved, and the $Sum$ is stored directly in the $M$-memristor. With SAPPI-2 we further improve the accuracy of the approximation and store the $Sum$ output in the $A$-memristor. Therefore only the $B$ input is preserved in this version.

\subsection{SAPPI-1}

In this algorithm, we utilize the property of IMPLY that the second operand is overwritten and the result of the operation is stored instead. For this, we use the work memristor to store the temporary results and apply the $A$ and $B$ inputs sequentially. With this the $Sum$ is stored in the $M$-memristor, keeping the original input states. We calculate the $C_{out}$ by applying $M\rightarrow C$ and store it in the $C$-memristor. So, we reduce the ER of $C_{out}$ to only $\frac{1}{8}$ where the only erroneous case is ``$A_{in}B_{in}C_{in}$'' = ``001''. The $Sum$ output has an error rate (ER) of $\frac{4}{8}$ which can be seen in the truth table in \Cref{table:truthtable}. The corresponding logical functions in boolean and IMPLY form for $Sum$ and $C_{out}$ are:
\begin{equation}
\centering
    Sum = B\rightarrow\overline{A} = \overline{AB}
\end{equation} \vspace{-1.5em}
\begin{equation}
\centering
    C_{out}=(B\rightarrow\overline{A})\rightarrow C= AB + C
\end{equation}

The exact procedure of SAPPI-1 is outlined in \Cref{table:stepssappi1}, beginning with a reset of the work memristor to logical '0' using a FALSE operation. In the second and third steps, the inputs are implied on the work memristor which stores the $Sum$ in the third step. In the fourth step, the $C_{out}$ is computed and saved in the $C$-memristor. This algorithm requires only $4n$ steps for an $n$-bit addition. 
When only SAPPI-1 adders are used, an $n$-bit addition would require $4n$ steps and $3n+1$ memristors.

\begin{table}[tb]
    \centering
    \caption{Truth table of proposed algorithms}
    \begin{tabular}{|c|c|c|c|c|c|c|c|c|}
    \hline
    \multicolumn{3}{|c|}{\textbf{Inputs}} & \multicolumn{2}{c|}{\textbf{Exact}} & \multicolumn{2}{c|}{\textbf{SAPPI-1}} & \multicolumn{2}{c|}{\textbf{SAPPI-2}} \\ \hline
    A\textsubscript{in} & B\textsubscript{in} & C\textsubscript{in} & Sum & C\textsubscript{out} & Sum & C\textsubscript{out} & Sum & C\textsubscript{out} \\ 
    \hline \hline 
    0 & 0 & 0 & 0 & 0 & 1 \xmark & 0 \cmark & 1 \xmark & 0 \cmark \\ 
    0 & 0 & 1 & 1 & 0 & 1 \cmark & 1 \xmark & 0 \xmark & 1 \xmark \\
    0 & 1 & 0 & 1 & 0 & 1 \cmark & 0 \cmark & 1 \cmark & 0 \cmark \\
    0 & 1 & 1 & 0 & 1 & 1 \xmark & 1 \cmark & 0 \cmark & 1 \cmark \\
    1 & 0 & 0 & 1 & 0 & 1 \cmark & 0 \cmark & 1 \cmark & 0 \cmark \\
    1 & 0 & 1 & 0 & 1 & 1 \xmark & 1 \cmark & 1 \xmark & 1 \cmark \\
    1 & 1 & 0 & 0 & 1 & 0 \cmark & 1 \cmark & 1 \xmark & 1 \cmark \\
    1 & 1 & 1 & 1 & 1 & 0 \xmark & 1 \cmark & 1 \cmark & 1 \cmark \\
    \hline
    \end{tabular}
    \label{table:truthtable}
\end{table}

\begin{table}[tb]

\caption{Computational steps of the SAPPI-1}
\centering
\begin{tabular}{c c l} 
 \hline
 Step & Operation & Equivalent logic  \\ 
 \hline\hline
 1 & $ M\textsubscript{1} = 0 $ & \text{FALSE}$(M\textsubscript{1})$ \\
 2 & $ A \rightarrow M\textsubscript{1} = M\textsubscript{1}' $ & \text{NOT(A)} \\
 3 & $ B \rightarrow M\textsubscript{1}' = M\textsubscript{1}'' $ & $ \textit{Sum} = B \rightarrow \overline{A} $ \\
 4 & $ M\textsubscript{1}'' \rightarrow C = C' $ & $\textit{C\textsubscript{out}} = (B \rightarrow \overline{A}) \rightarrow C $ \\
 \hline
\end{tabular}
\label{table:stepssappi1}
\vspace{-2mm}
\end{table}

\subsection{SAPPI-2}
In SAPPI-2, the approach also involves sequentially implicating the inputs into the work memristor to store temporary results. However, instead of storing the $Sum$ result in the work memristor as done in SAPPI-1, the $Sum$ is saved in the $A$-memristor using an additional step. This modification aligns with the exact serial algorithm from~\cite{Rohani2017MemFA} and ensures compatibility. As with SAPPI-1, the ER for $C_{out}$ is $\frac{1}{8}$. Although the $Sum$ output has an ER of $\frac{4}{8}$, two error placements have been altered as shown in~\Cref{table:truthtable}. The logical equations for SAPPI-2 are:

\begin{equation}
\centering
    Sum = ((B \rightarrow \overline{A}) \rightarrow C) \rightarrow A = \overline{AB + C} + A 
\end{equation}\vspace{-1.5em}
\begin{equation}
\centering
    C_{out}=  (B \rightarrow \overline{A}) \rightarrow C = AB + C
\end{equation}

A notable aspect is that a single erroneous bit in the $Sum$ now mirrors the error placement in $C_{out}$, meaning both $Sum$ and $C_{out}$ are swapped in this scenario. This results in a reduced overall error, as indicated by improved error metrics. The exact procedure for SAPPI-2 is described in \Cref{table:stepssappi2}, which is similar to SAPPI-1 except for an additional fifth step where the $Sum$ is stored in the $A$-memristor. This algorithm requires $5n$ steps and uses $2n+2$ memristors when it is applied $n$ times.

\begin{table}[tb]
\caption{Computational steps of the SAPPI-2}
\centering
\begin{tabular}{c c l} 
 \hline
 Step & Operation & Equivalent logic  \\ 
 \hline\hline
 1 & $ M\textsubscript{1} = 0 $ & \text{FALSE}$(M\textsubscript{1})$ \\
 2 & $ A \rightarrow M\textsubscript{1} = M\textsubscript{1}' $ & \text{NOT(A)} \\
 3 & $ B \rightarrow M\textsubscript{1}' = M\textsubscript{1}'' $ & $ B \rightarrow \text{NOT(A)} $ \\
 4 & $ M\textsubscript{1}'' \rightarrow C = C' $ & $\textit{C\textsubscript{out}} = (B \rightarrow \overline{A}) \rightarrow C $ \\
 5 & $ C' \rightarrow A = A' $ & $\textit{Sum} = ((B \rightarrow \overline{A}) \rightarrow C) \rightarrow A $ \\
 \hline
\end{tabular}
\label{table:stepssappi2}
\end{table}

\section{Circuit-level Simulation and Error Metrics}

\subsection{Simulation Setup}
We simulated the proposed approximated full adders via LT-SPICE to verify their functionality. We used the Voltage Threshold Adaptive Memristor (VTEAM) model implemented in SPICE~\cite{Jungwirth2018SPICE} with the parameters fitted to a discrete Knowm memristor~\cite{Jungwirth2018SPICE, knowm}, shown in \Cref{table:VTEAMmodel}. We have to note that similar to the differences in discrete and integrated CMOS devices, discrete memristors lead to slower operations and increased power consumption. It is therefore reasonable to assume that integrated memristors have a significantly better performance. We chose the IMPLY-specific parameters similar to \Cref{table:IMPLYparameter} to allow for a fair comparison to SoA papers such as~\cite{Fatemieh2023AFAIP, Asgari2024SAFAN, Seiler2023SSAxA}. 
Since real memristors are non-ideal devices, we included the deviation of $R_{on}$ and $R_{off}$ in our experiments as the resistive deviation is one of the most important behaviors that has to be expected. We simulated all eight input combinations with different deviations of up to $\pm 30\%$ which can be seen as the shaded area in \Cref{fig:simulations_of_SAPPI1} and \Cref{fig:simulations_of_SAPPI2}. 

\begin{table}[tb]
\centering
\caption{IMPLY parameter values \cite{Fatemieh2023AFAIP}}
\begin{tabular}{|c|c|c|c|c|c|}
\hline
Parameter & $V_{SET}$ & $V_{RESET}$ & $V_{COND}$ & $R_{G}$ & $t_{pulse}$ \\
\hline
Value & 1 V & -1 V & 900 mV & 40 k$\Omega$ & 30 $\mu$s \\
\hline
\end{tabular}
\label{table:IMPLYparameter}
\end{table}

\begin{table}[tb]
\centering
\caption{VTEAM model and setup values \cite{Fatemieh2023AFAIP}}
\resizebox{1\columnwidth}{!}{
\begin{tabular}{|c|c|c|c|c|c|c|c|}
\hline
Parameter & $v_{off}$ & $v_{on}$ & $\alpha_{off}$ & $\alpha_{on}$ & $R_{off}$ & $R_{on}$ \\
\hline
Value & 0.7 V & -10 mV & 3 & 3 & 1 M$\Omega$ & 10 k$\Omega$ \\
\hline
$k_{on}$ & $k_{off}$ & $w_{off}$ & $w_{on}$ & $w_{C}$ & $a_{off}$ & $a_{on}$ \\
\hline
-0.5 nm/s & 1 cm/s & 0 nm & 3 nm & 107 pm & 3 nm & 0 nm \\
\hline
\end{tabular}
}
\label{table:VTEAMmodel}
\end{table}

\subsection{Simulation Results}

\begin{figure}[!bt] 
\centering
\subfloat[``AinBinCin''=``100'' with correct output]{\includegraphics[width=1\columnwidth]{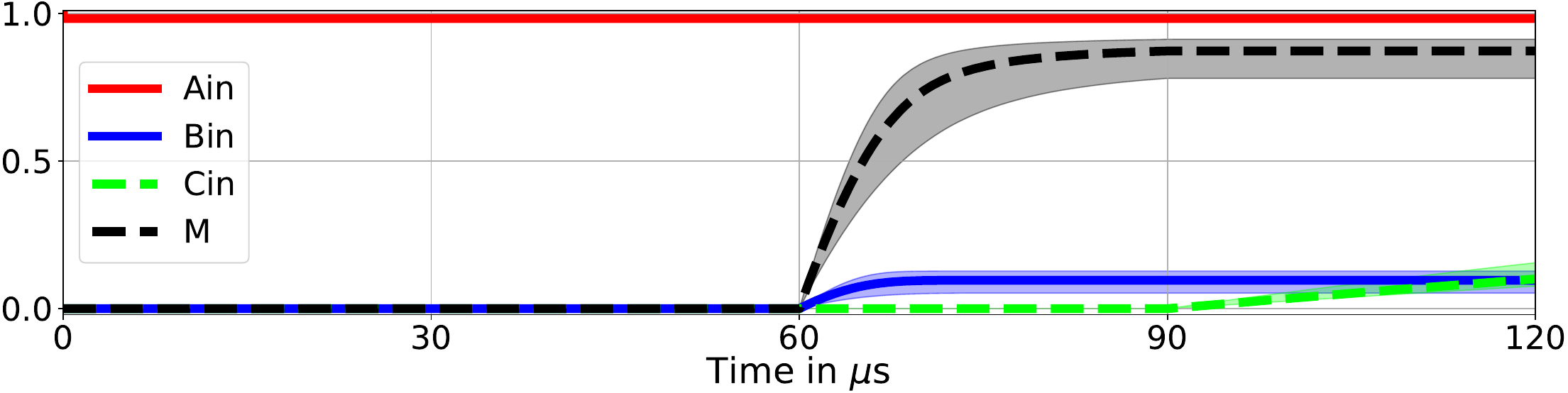}}
\vspace{-0.5mm}
\subfloat[``AinBinCin''=``001'' with approximated (erroneous by design) output.]{\includegraphics[width=1\columnwidth]{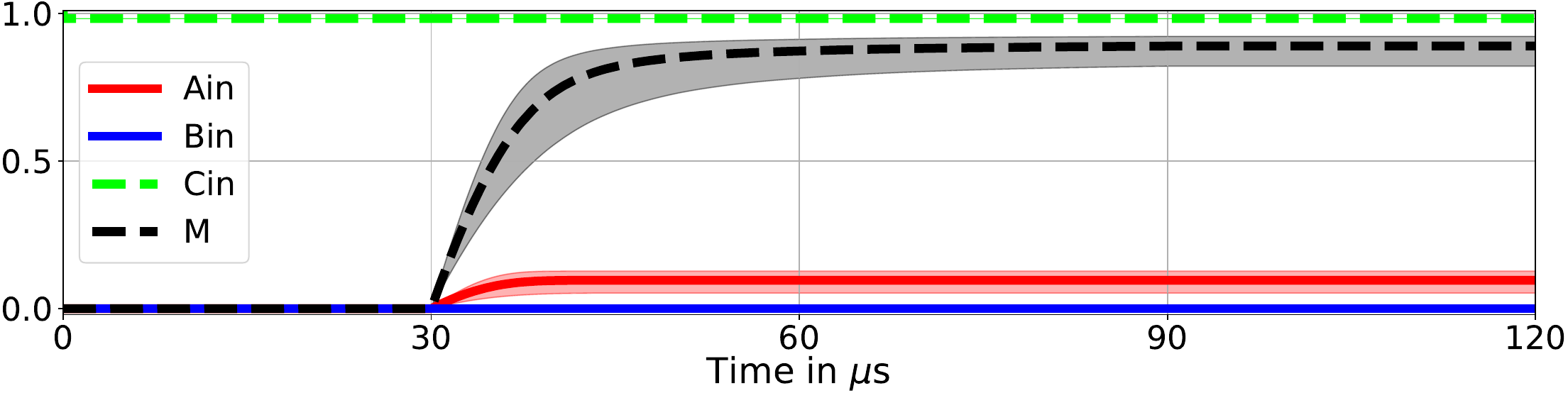}}
\caption{Two example simulations of SAPPI-1, illustrating the resistive deviation of $\pm 30\%$ as shaded areas.}
\label{fig:simulations_of_SAPPI1}
\end{figure}

\begin{figure}[!bt] 
\centering
\subfloat[``AinBinCin''=``010''  with correct output]{\includegraphics[width=1\columnwidth]{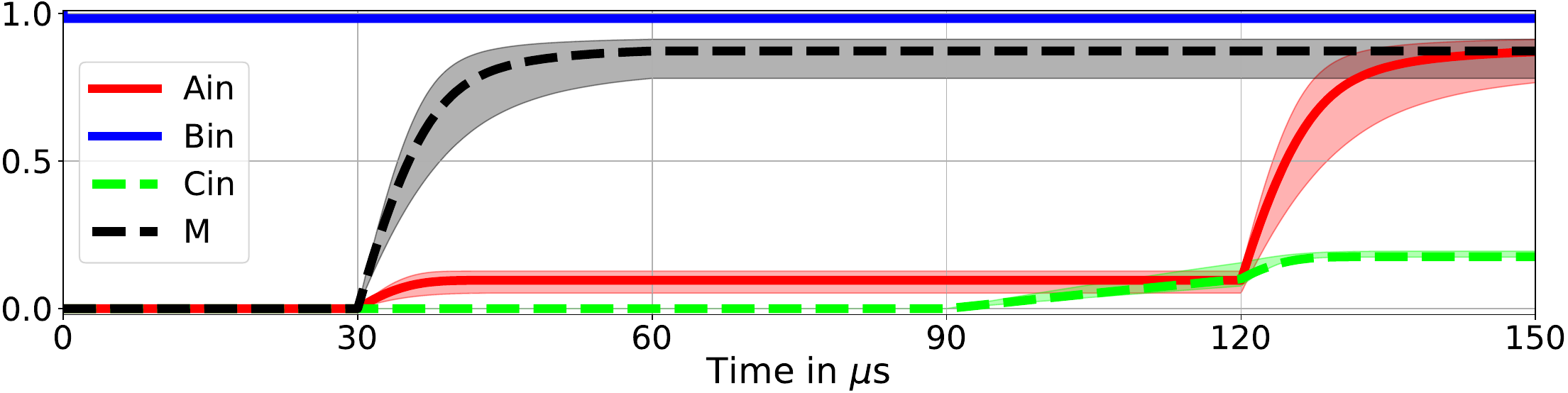}}
\vspace{-0.5mm}
\subfloat[``AinBinCin''=``000'' with approximated (erroneous by design) output.]{\includegraphics[width=1\columnwidth]{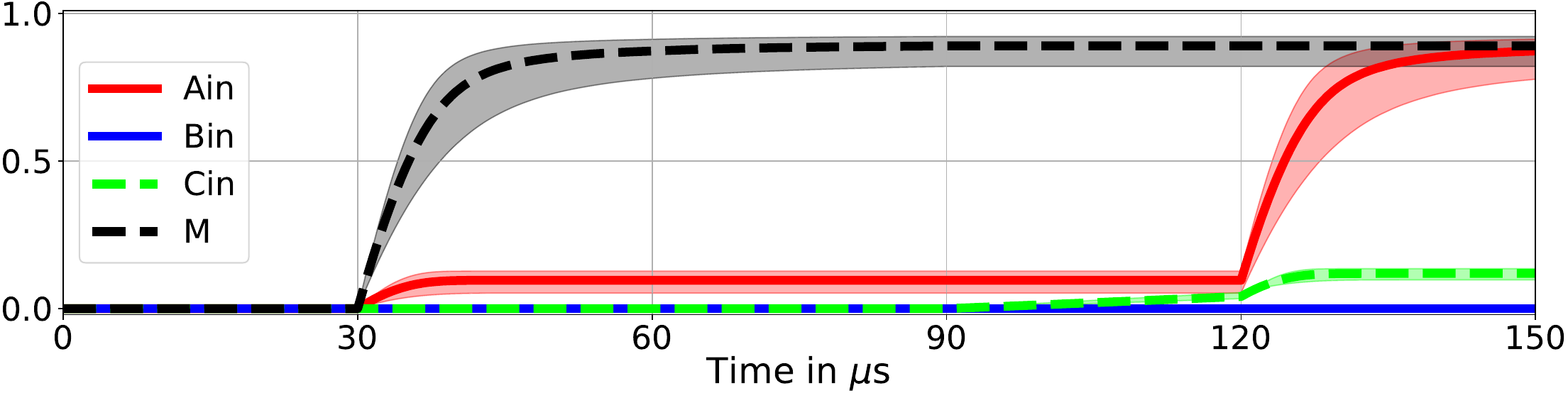}}
\caption{Two example simulations of SAPPI-2, illustrating the resistive deviation of $\pm 30\%$ as shaded areas.}
\label{fig:simulations_of_SAPPI2}

\end{figure}

For both presented algorithms, we evaluated all eight input combinations with deviations of up to $\pm 30\%$. All combinations agree with the intended behavior as $Sum$ and $C_{out}$ match the corresponding truth table. Each step in the algorithms corresponds to $30\mu s$ in the simulations. We arbitrarily illustrated one exact and one incorrect by-design case for SAPPI-1 in \Cref{fig:simulations_of_SAPPI1}. The $Sum$ output is saved in the $M_1$-memristor at 60$\mu s$-90$\mu s$ which corresponds to the third step. In the fourth step from 90$\mu s$-120$\mu s$, the $C_{out}$ is stored in the $C$-memristor. The waveforms for SAPPI-2 are shown in \Cref{fig:simulations_of_SAPPI2} where the $C_{out}$ is stored in the $C$-memristor at 90$\mu s$-120$\mu s$ which corresponds to step four. For this algorithm, the $Sum$ is stored in the $A$-memristor at the fifth step (120$\mu s$-150$\mu s$).

\subsection{Error Metrics}

To quantify the erroneous behavior of the presented approximated adders, we used common error metrics such as MED, NMED and MRED to efficiently evaluate the accuracy. More details on these metrics can be found in~\cite{Jiang2017AxReview, Jiang2020AAC}. We evaluated all input cases for an 8-bit RCA that was partially approximated by the presented adders. The results are shown in \Cref{table:error_metrics_8bit_rca} where we can see that both presented adders have very similar metrics until an approximation degree of 4/8. With more approximated adders, the accuracy of SAPPI-1 is getting increasingly worse compared to SAPPI-2. This indicates that the intentional error at the case ``A$_{in}$B$_{in}$C$_{in}$" = ``001" leads to a non-mitigated error propagation for SAPPI-1. In SAPPI-2 this case is erroneous for both $Sum$ and $C_{out}$ leading to a mitigation of error.

\begin{table}[tb]
\caption{Error Metrics of the presented algorithms in an 8-bit approximated full adder (Ax FA) with different approximation degrees}
\resizebox{1\columnwidth}{!}{
\begin{tabular}{|c|c|c|c|c|c|c|}
\hline
\multirow{2}{*}{\textbf{Ax FA}} &\multicolumn{3}{c|}{\textbf{SAPPI-1}} & \multicolumn{3}{c|}{\textbf{SAPPI-2}}\\ \cline{2-7}
 & MED & NMED & MRED & MED & NMED & MRED\\ \hline
Ax FA: 1/8   & 0.2500 & 0.0004 & 0.0013 & 0.5000 & 0.0009 & 0.0027 \\
Ax FA: 2/8   & 1.2500 & 0.0024 & 0.0069 & 1.5000 & 0.0029 & 0.0082 \\
Ax FA: 3/8   & 3.5312 & 0.0069 & 0.0197 & 3.5000 & 0.0068 & 0.0194 \\
Ax FA: 4/8   & 8.6250 & 0.0169 & 0.0492 & 7.5000 & 0.0147 & 0.0423 \\
Ax FA: 5/8   & 19.6347 & 0.0385 & 0.1156 & 15.5000 & 0.0303 & 0.0896 \\
Ax FA: 8/8   & 191.0572 & 0.3746 & 1.4026 & 127.5000 & 0.2500 & 0.8841 \\
\hline
\end{tabular}
}
\label{table:error_metrics_8bit_rca}
\end{table}

\section{Circuit-level Comparison}
\label{Circuit-level Comparison}

In this section, we compare the relevant circuit-level criteria with exact~\cite{Rohani2017MemFA, Karimi2018MemFA} and approximated~\cite{Fatemieh2023AFAIP, Asgari2024SAFAN} adders that are based on the serial topology. We do not compare to other algorithms in other topologies~\cite{Seiler2023SSAxA} in this work as they are built for different optimization goals. An overview of the different criteria compared to SoA algorithms is shown in \Cref{table:circuit_Level_Comparison}. 
Since IMPLY-based approaches are algorithm-based, the approximation level can be reconfigured for each process or even during the run. In the following sections, we just compare with an approximation degree of 4/8 as it is the highest approximation degree that leads to acceptable results on application-level.

\subsection{Energy Consumption}

We recreated the energy consumption of the exact serial algorithm from~\cite{Rohani2017MemFA} with our specified simulation parameters in LT-SPICE as our partially approximated RCA uses this algorithm at the higher bits. The energy consumption for the presented and recreated adders was evaluated as the mean of all different input cases. The total energy consumption of such an RCA with $k$ approximated out of $n$ total bits is:
\begin{equation}
{E}_{\textit{SAPPI-1}}(n, k)  = 0.7980k + 4.8250(n-k)
\label{eqnEN}
\end{equation}\vspace{-1.5em}
\begin{equation}
{E}_{\textit{SAPPI-2}}(n, k) = 1.0919k + 4.8250(n-k) 
\label{eqnEN2}
\end{equation}

SAPPI-1 is more energy efficient than SAPPI-2 and requires 5\% less energy.
Compared to exact adders the presented algorithms reduce the required energy by 39\%-42\%. Even to the already optimized approximated adders from~\cite{Asgari2024SAFAN, Fatemieh2023AFAIP} our proposed methods improve the energy consumption by 9\%-13\%, rendering our algorithms more efficient than the SoA.

\subsection{Number of Steps}

Another important criterion on the circuit level is the number of steps that are required per bit. SAPPI-1 and SAPPI-2 require only 4 and 5 steps per bit. When we now include them as the lower bits in an RCA where the higher bits are again calculated by the exact adder from~\cite{Rohani2017MemFA}, the total number of steps with $k$ approximated and $n-k$ exact adders are:
\begin{equation}
S_\text{SAPPI-1}(n, k)  =4k + 22(n - k)
\label{eqn3}
\end{equation}\vspace{-1.5em}
\begin{equation}
S_\text{SAPPI-2}(n, k)  = 5k + 22(n - k)
\label{eqn4}
\end{equation}

As our approach requires 17-18 steps per bit less than the exact adder from~\cite{Rohani2017MemFA}, up to 39\%-41\% steps are saved. Compared to the fastest approximated adder from~\cite{Asgari2024SAFAN, Fatemieh2023AFAIP} which is SAFAN, our adders require 7\%-10\% fewer steps. 

\begin{table*}[tb]

\centering
    \caption{Circuit-Level Comparison to the SoA full adder}
\resizebox{\textwidth}{!}{    
\begin{tabular}{|c|c|c|c|c|c|c|c|c|c|}
\hline
 \textbf{Full adder} & \multicolumn{3}{c|}{\textbf{Energy consumption (nJ)}} & \multicolumn{3}{c|}{\textbf{No. of steps}} & \multicolumn{2}{c|}{\textbf{No. of memristors}} \\
 \cline{2-9}
  & \textbf{n, k} & \textbf{n=8-bit, k=4} & \textbf{Imp. to ~\cite{Rohani2017MemFA}} & \textbf{n, k} & \textbf{n=8-bit, k=4} & \textbf{Imp. to ~\cite{Rohani2017MemFA}} & \textbf{n, k} & \textbf{n=8-bit, k=4} \\
 \hline \hline
 Exact 1 \cite{Rohani2017MemFA} & 4.8250n & 38.6000 & - & 22n & 176 & - & 2n+3 & 19 \\
 Exact 2 \cite{Karimi2018MemFA} & 4.0772n & 32.6176 & 15\% & 23n & 184 & -4\% & 2n+3 & 19 \\
 \hline
 SIAFA1,3 \cite{Fatemieh2023AFAIP} & 1.7090k + 4.8250(n-k) & 26.1360 & 32\% & 8k + 22(n-k) & 120 & 32\% & 2n+3 & 19 \\
 SIAFA2 \cite{Fatemieh2023AFAIP} & 2.5131k + 4.8250(n-k) & 29.3524 & 24\% & 10k + 22(n-k) & 128 & 27\% & 2n+3 & 19 \\
 SIAFA4 \cite{Fatemieh2023AFAIP} & 1.7066k + 4.8250(n-k) & 26.1264 & 32\% & 8k + 22(n-k) & 120 & 32\% & 2n+3 & 19 \\
 SAFAN \cite{Asgari2024SAFAN} & 1.6628k + 4.8250(n-k) & 25.9512 & 33\% & 7k + 22(n-k) & 116 & 34\% & 2n+3 & 19 \\
 \hline
 \textbf{SAPPI-1} & \textbf{0.7980k + 4.8250(n-k)} & \textbf{22.4920} & \textbf{42\%} & \textbf{4k + 22(n-k)} & \textbf{104} & \textbf{41\%} & 2n+k+3 & 23 \\
 \textbf{SAPPI-2} & 1.0919k + 4.8250(n-k) & 23.6676 & 39\% & 5k + 22(n-k) & 108 & 39\% & 2n+3 & 19 \\
 \hline
    \end{tabular}
    }
    \label{table:circuit_Level_Comparison}
\begin{tablenotes}
      \small
      
      \item We have simulated the circuits from~\cite{Fatemieh2023AFAIP, Asgari2024SAFAN,Rohani2017MemFA, Karimi2018MemFA} similar to ours and obtained different numbers to the reported values \\ from~\cite{Asgari2024SAFAN, Fatemieh2023AFAIP} and are not sure why. So to allow for a fair comparison, we are reporting our own simulated results.
    \end{tablenotes}
\end{table*}

\subsection{Area}

For SAPPI-2 and SoA approximations from~\cite{Fatemieh2023AFAIP, Asgari2024SAFAN} the work memristors are reused for each bit. So the number of memristors is bounded by the exact serial algorithm from~\cite{Rohani2017MemFA} which means they require $2n+3$ memristors for $n$-bit. In SAPPI-1 the $Sum$ result is stored in the work memristor and the inputs are preserved. As the work memristors cannot be reused, additional $k$ memristors are required. So depending on the approximation degree, SAPPI-1 needs $2n+k+3$ memristors for a $n$-bit addition. This does not necessarily mean it is more area-inefficient as it preserves the inputs of the adder, compared to other approaches.

\section{Application in Image Processing}

\begin{figure}[tb]
  \centering
  
  \subfloat[\hspace{-1mm}cameraman]{\includegraphics[width=0.19\columnwidth]{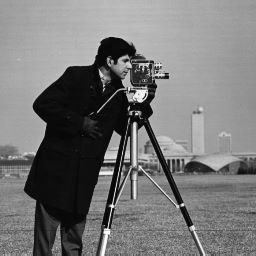}}\hspace{0.01mm}
  \subfloat[rice]{\includegraphics[width=0.19\columnwidth]{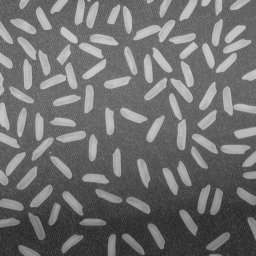}}\hspace{0.01mm}
  \subfloat[exact]{\includegraphics[width=0.19\columnwidth]{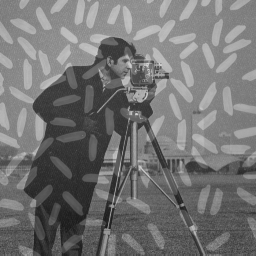}}\hspace{0.01mm}
  \subfloat[SAPPI-1]{\includegraphics[width=0.19\columnwidth]{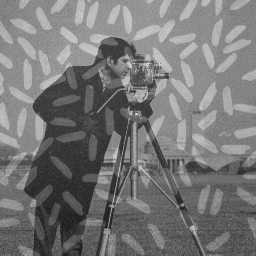}}\hspace{0.01mm}
  \subfloat[SAPPI-2]{\includegraphics[width=0.19\columnwidth]{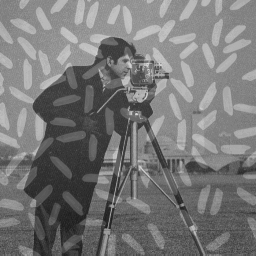}}%
    
  \vspace{-1em}

  \subfloat[toys]{\includegraphics[width=0.24\columnwidth]{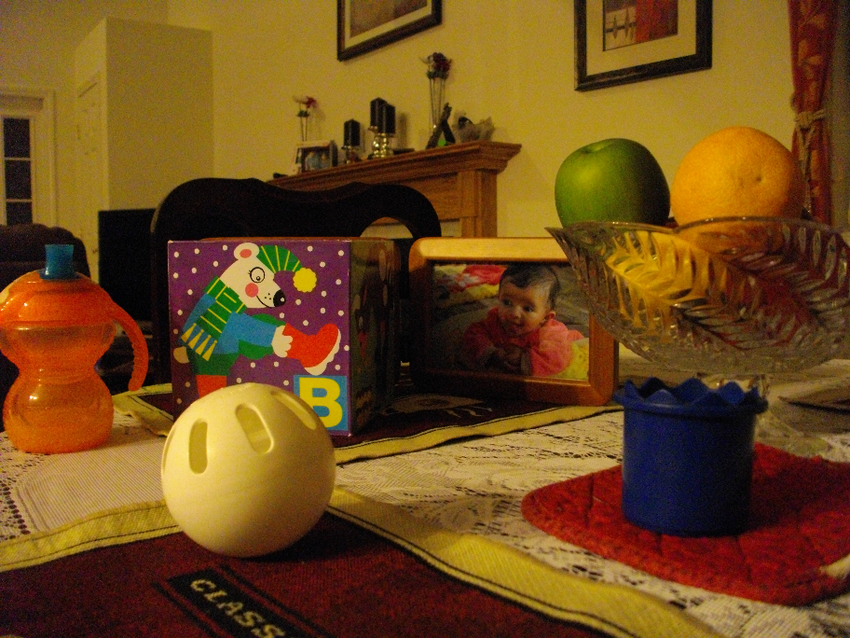}}\hspace{0.1mm}
  \subfloat[exact]{\includegraphics[width=0.24\columnwidth]{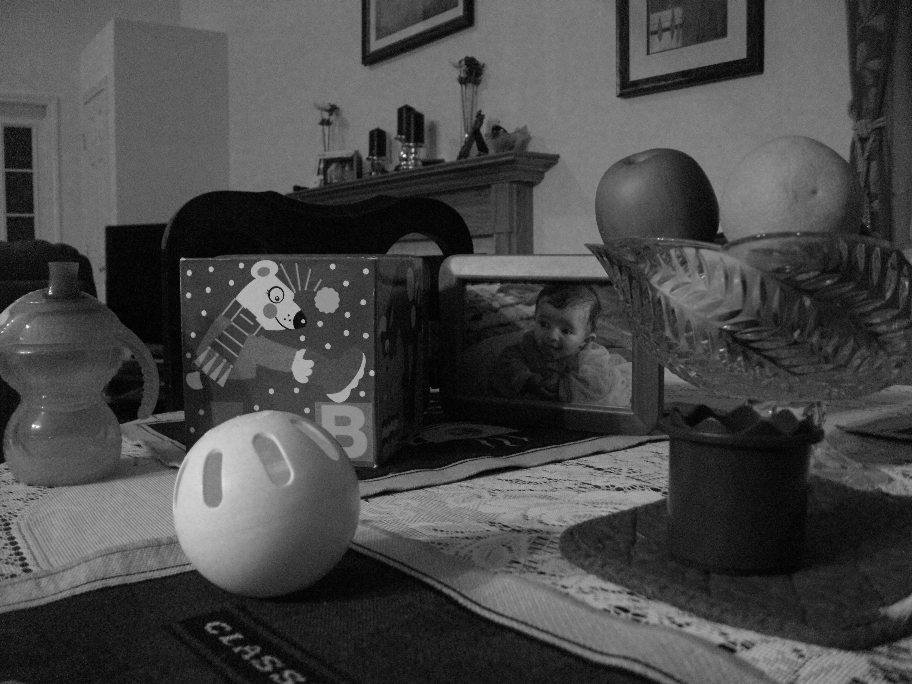}}\hspace{0.1mm}
  \subfloat[SAPPI-1]{\includegraphics[width=0.24\columnwidth]{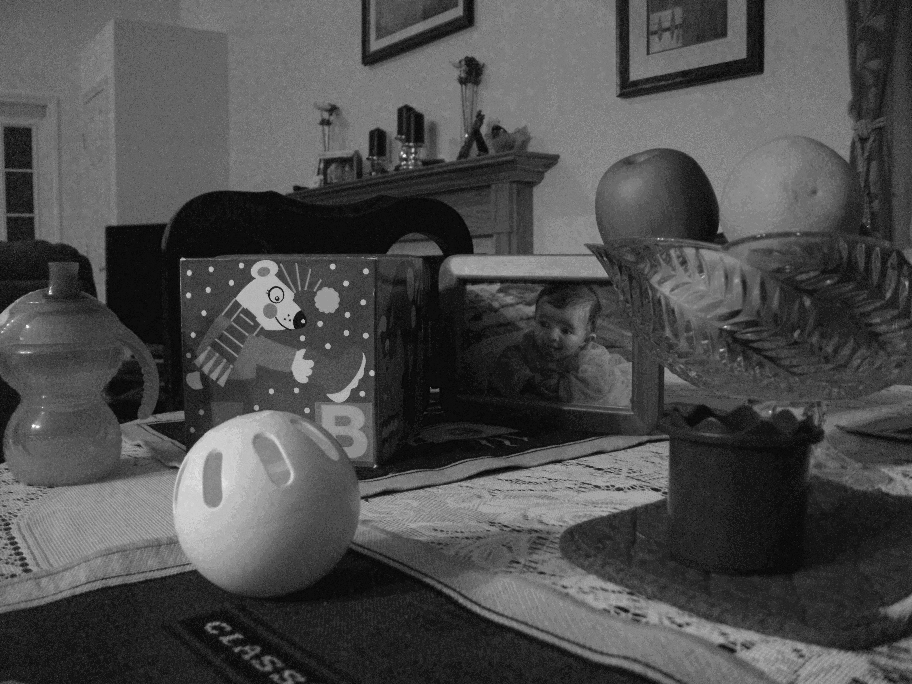}}\hspace{0.1mm}
    \subfloat[SAPPI-2]{\includegraphics[width=0.24\columnwidth]{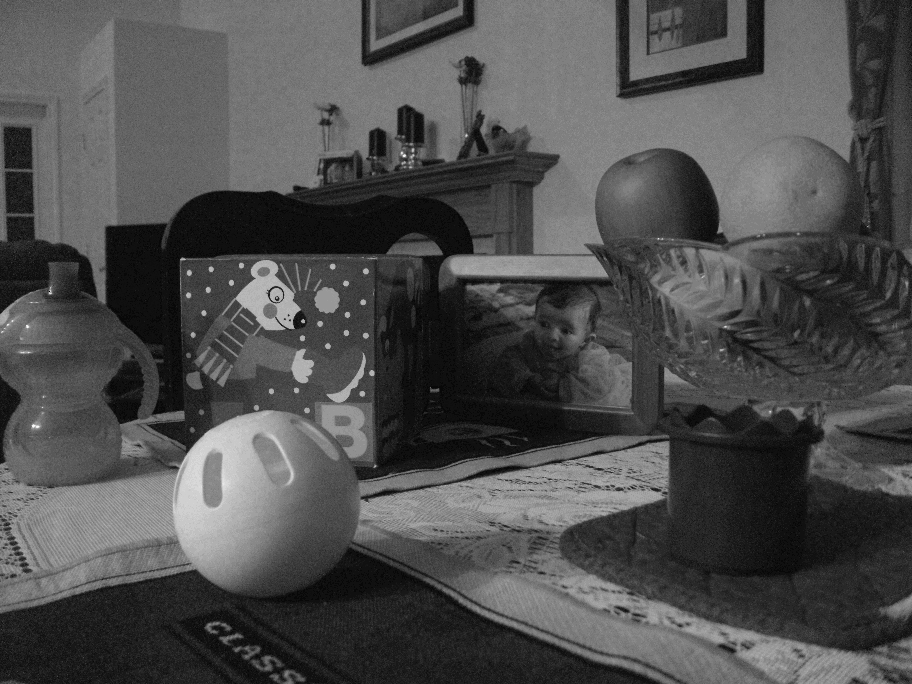}}%
    
  \vspace{-1em}

  \subfloat[boats]{\includegraphics[width=0.24\columnwidth]{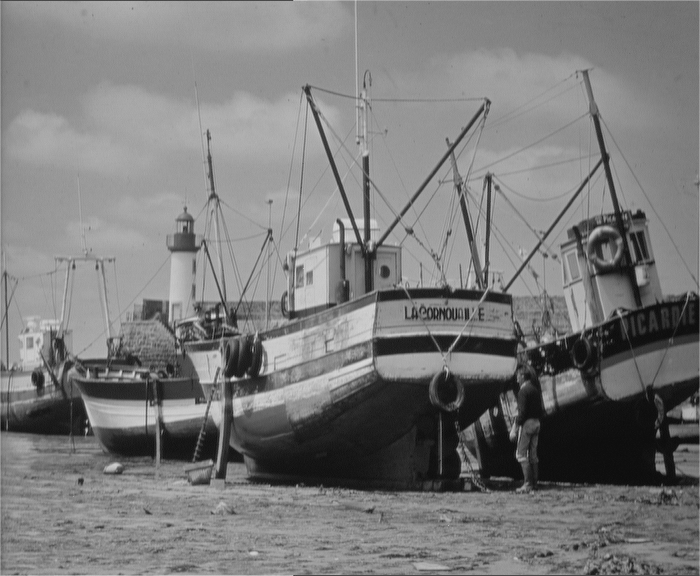}}\hspace{0.1mm}
    \subfloat[exact]{\includegraphics[width=0.24\columnwidth]{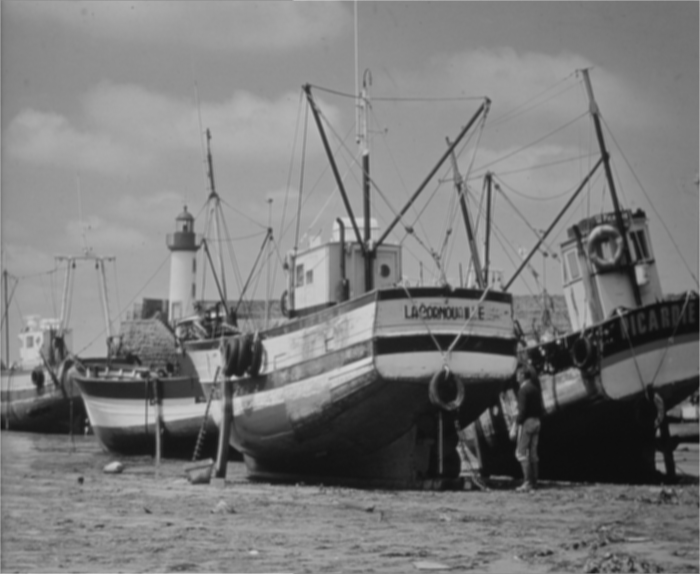}}\hspace{0.1mm}
  \subfloat[SAPPI-1]{\includegraphics[width=0.24\columnwidth]{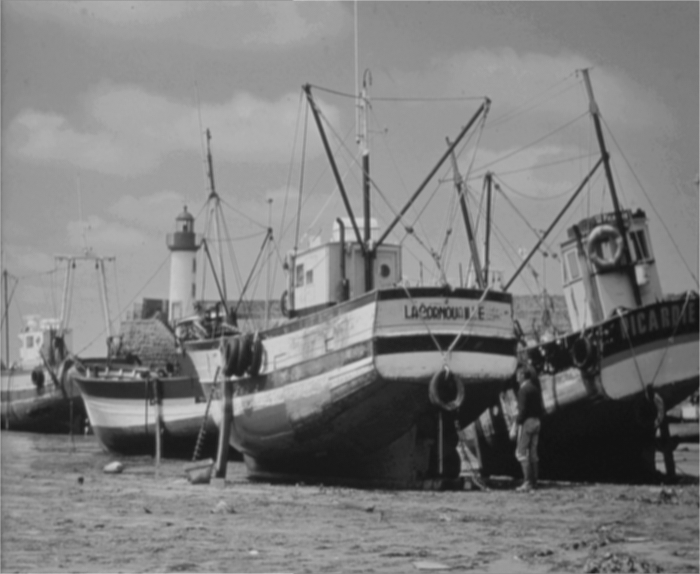}}\hspace{0.1mm}
        \subfloat[SAPPI-2]{\includegraphics[width=0.24\columnwidth]{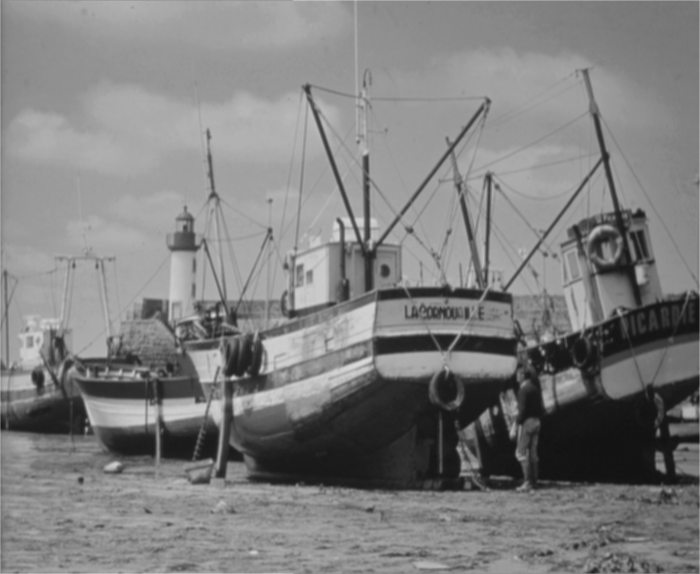}}%
        \caption{Results of different image processing applications with 4/8 Ax FA: image addition (top), grayscale conversion (middle), and with 8/20 Ax FA: Gaussian blurring (bottom).}
    \label{fig:combined}
 \end{figure}

Image processing is an error-resilient application widely used in computer vision, robotics, industrial systems and media \cite{Khaleqi2021ImpreciseMultipliers}. We simulated three common applications with RCA that partially consist of our approximated full adders. We evaluated different approximation degrees and determined the quality metrics PSNR and MSSIM which are represented in \Cref{table:QualityMetrics}. In the literature, a PSNR of over 30dB is deemed acceptable quality. We presented the resulting images with half of the adders approximated in~\Cref{fig:combined}. In \Cref{Section6.2} we analyze possible gains of our approach on application-level and compare to SoA approximations.

\subsection{Example Applications}
\subsubsection{Image Addition}

Image addition is used for masking and enhancing images \cite{Gupta2013DSP}. We simulated this by adding the corresponding pixels of two images via an RCA and halving the result. We evaluated two 8-bit standard images with varying approximation degrees. Both of the presented algorithms achieved a PSNR of over 30dB with up to 4/8 adders approximated which demonstrates acceptable image quality.

\begin{table*}[tb]
    \centering
    \caption{Application-level gains to exact serial algorithm for different image processing applications.}
    \label{tab:my_label}
    \resizebox{\textwidth}{!}{%
    \begin{tabular}{|c|c|c|c|c|c|c|c|}
    \hline
    \multirow{2}{*}{\textbf{Ax algorithm}} & \multicolumn{2}{c|}{\textbf{Image addition (256$\times$256 8-bit Images)}} & \multicolumn{2}{c|}{\textbf{Grayscale conversion (684$\times$912$\times$3 8-bit Image)}} & \multicolumn{2}{c|}{\textbf{Gaussian blurring (576$\times$700 8-bit Image)}} \\ \cline{2-7}
    & Energy saved (mJ) & Steps saved (million) & Energy saved (mJ) & Steps saved (million) & Energy saved (mJ) & Steps saved (million) \\ \hline
    \textbf{SAPPI-1} (4/8 Ax FA) &1.0557 &18.8744 &20.0966 &359.3134 &580.8332 &2596.2250 \\ \hline
    \textbf{SAPPI-2} (4/8 Ax FA) &0.9786 &17.8258 &18.6299 &339.3516 &538.4426 &2451.9902 \\ \hline
    \end{tabular}
    }
\end{table*}

\begin{table}[bt]
\caption{Quality metrics of image processing applications}
\resizebox{\columnwidth}{!}{%
\begin{tabular}{|lcccccc|}
\hline
\multicolumn{1}{|c|}{\multirow{2}{*}{\textbf{\begin{tabular}[c]{@{}c@{}}Ax\\ algorithm\end{tabular}}}} &
  \multicolumn{2}{c|}{\textbf{\begin{tabular}[c]{@{}c@{}}Image \\ addition\end{tabular}}} &
  \multicolumn{2}{c|}{\textbf{\begin{tabular}[c]{@{}c@{}}Grayscale \\ filter\end{tabular}}} &
  \multicolumn{2}{c|}{\textbf{\begin{tabular}[c]{@{}c@{}}Gaussian \\ blurring\end{tabular}}} \\ \cline{2-7} 
\multicolumn{1}{|c|}{} &
  \multicolumn{1}{c|}{\begin{tabular}[c]{@{}c@{}}PSNR\\ (dB)\end{tabular}} &
  \multicolumn{1}{c|}{MSSIM} &
  \multicolumn{1}{c|}{\begin{tabular}[c]{@{}c@{}}PSNR\\ (dB)\end{tabular}} &
  \multicolumn{1}{c|}{MSSIM} &
  \multicolumn{1}{c|}{\begin{tabular}[c]{@{}c@{}}PSNR\\ (dB)\end{tabular}} &
  MSSIM \\ \hline
\multicolumn{5}{|c|}{Approximation degree: 1/8 Ax FA} & \multicolumn{2}{c|}{2/20 Ax FA} \\ \hline
\multicolumn{1}{|l|}{SAPPI-1} &
  \multicolumn{1}{c|}{54.10} &
  \multicolumn{1}{c|}{0.9992} &
  \multicolumn{1}{c|}{52.34} &
  \multicolumn{1}{c|}{0.9982} &
  \multicolumn{1}{c|}{88.98} &
   1.0000 \\ \hline
\multicolumn{1}{|l|}{SAPPI-2} &
  \multicolumn{1}{c|}{51.12} &
  \multicolumn{1}{c|}{0.9989} &
  \multicolumn{1}{c|}{49.43} &
  \multicolumn{1}{c|}{0.9982} &
   \multicolumn{1}{c|}{79.12} &
   1.0000 \\ \hline
\multicolumn{5}{|c|}{Approximation degree: 2/8 Ax FA} & \multicolumn{2}{c|}{4/20 Ax FA} \\ \hline
\multicolumn{1}{|l|}{SAPPI-1} &
  \multicolumn{1}{c|}{48.10} &
  \multicolumn{1}{c|}{0.9974} &
  \multicolumn{1}{c|}{46.08} &
  \multicolumn{1}{c|}{0.9949} &
 \multicolumn{1}{c|}{72.82} &
   1.0000 \\ \hline
\multicolumn{1}{|l|}{SAPPI-2} &
  \multicolumn{1}{c|}{46.34} &
  \multicolumn{1}{c|}{0.9978} &
  \multicolumn{1}{c|}{43.71} &
  \multicolumn{1}{c|}{0.9949} &
  \multicolumn{1}{c|}{65.53} &
   1.0000 \\ \hline
\multicolumn{5}{|c|}{Approximation degree: 3/8 Ax FA} & \multicolumn{2}{c|}{6/20 Ax FA} \\ \hline
\multicolumn{1}{|l|}{SAPPI-1} &
  \multicolumn{1}{c|}{40.51} &
  \multicolumn{1}{c|}{0.9866} &
  \multicolumn{1}{c|}{38.95} &
  \multicolumn{1}{c|}{0.9758} &
 \multicolumn{1}{c|}{54.08} &
   0.9998 \\ \hline
\multicolumn{1}{|l|}{SAPPI-2} &
  \multicolumn{1}{c|}{40.70} &
  \multicolumn{1}{c|}{0.9937} &
  \multicolumn{1}{c|}{37.84} &
  \multicolumn{1}{c|}{0.9827} &
   \multicolumn{1}{c|}{48.75} &
   0.9998 \\ \hline
\multicolumn{5}{|c|}{Approximation degree: 4/8 Ax FA} & \multicolumn{2}{c|}{8/20 Ax FA} \\ \hline
\multicolumn{1}{|l|}{SAPPI-1} &
  \multicolumn{1}{c|}{33.42} &
  \multicolumn{1}{c|}{0.9420} &
  \multicolumn{1}{c|}{31.91} &
  \multicolumn{1}{c|}{0.8936} &
   \multicolumn{1}{c|}{35.46} &
   0.9893 \\ \hline
\multicolumn{1}{|l|}{SAPPI-2} &
  \multicolumn{1}{c|}{35.01} &
  \multicolumn{1}{c|}{0.9800} &
  \multicolumn{1}{c|}{31.76} &
  \multicolumn{1}{c|}{0.9378} &
   \multicolumn{1}{c|}{33.57} &
   0.9942 \\ \hline
\multicolumn{5}{|c|}{Approximation degree: 5/8 Ax FA} & \multicolumn{2}{c|}{10/20 Ax FA} \\ \hline
\multicolumn{1}{|l|}{SAPPI-1} &
  \multicolumn{1}{c|}{26.03} &
  \multicolumn{1}{c|}{0.8193} &
  \multicolumn{1}{c|}{24.65} &
  \multicolumn{1}{c|}{0.6764} &
   \multicolumn{1}{c|}{20.33} &
   0.9092 \\ \hline
\multicolumn{1}{|l|}{SAPPI-2} &
  \multicolumn{1}{c|}{28.52} &
  \multicolumn{1}{c|}{0.9408} &
  \multicolumn{1}{c|}{25.28} &
  \multicolumn{1}{c|}{0.8004} &
  \multicolumn{1}{c|}{19.69} &
   0.9331 \\ \hline
\end{tabular}%
}\label{table:QualityMetrics}
\vspace{-2mm}
\end{table}

\subsubsection{Gray Scale Conversion}

The grayscale conversion is a standard pre-processing procedure for different computer vision and ML tasks. It converts an RGB image to grayscale by averaging the red, green and blue values of each pixel. To perform grayscaling the individual color values are added together in two iterations and the result is then divided by three. We simulated with varying approximation degrees and achieved acceptable quality (PSNR$>$30dB) for both presented algorithms with up to 4/8 adders approximated.

\subsubsection{Gaussian Smoothing} 
To showcase the applicability of our approximated adders on more complex tasks, we evaluated Gaussian smoothing as it is often used for denoising and as a preprocessing step in various computer vision tasks. We implemented this by using a partially approximated 20-bit RCA which we embedded in a shift-and-add multiplier to apply the 3$\times$3 kernel from~\cite{Amirafshar2023CDM} over the whole image. 
We simulated the Gaussian smoothing with varying approximation degrees and achieved acceptable PSNR over 30dB for both presented algorithms with up to 8/20 adders approximated.

\subsection{Application-Level Gains and Comparison}
\label{Section6.2}

The presented approaches exhibit substantial gains in energy efficiency and required cycles over the exact algorithm~\cite{Rohani2017MemFA}. Table \ref{tab:my_label} presents an overview of the possible gains for the previously covered image processing tasks. Our algorithms reduce the number of steps by $39\%-41\%$ and the energy consumption by $39\%-42\%$. In image addition we can save up to 1.1 mJ of energy and up to 19 million steps. In the grayscale conversion example, up to 20 mJ of energy and 359 million steps can be saved when our approach is applied. For Gaussian blurring of a 576$\times$700 8-bit image, these gains sum up to 581 mJ of energy and 2596 million steps.
Compared to SoA approximations our approach is not able to reach the 30dB threshold of PSNR with 5/8 approximated adders. But with four adders approximated, our adders reduce the number of steps by $7\%-10\%$ and improve the energy efficiency by $9\%-13\%$ compared to the most efficient SoA approximation (SAFAN).

\section{Application in Machine Learning}

\begin{figure}
    \centering
    \includegraphics[width=\columnwidth]{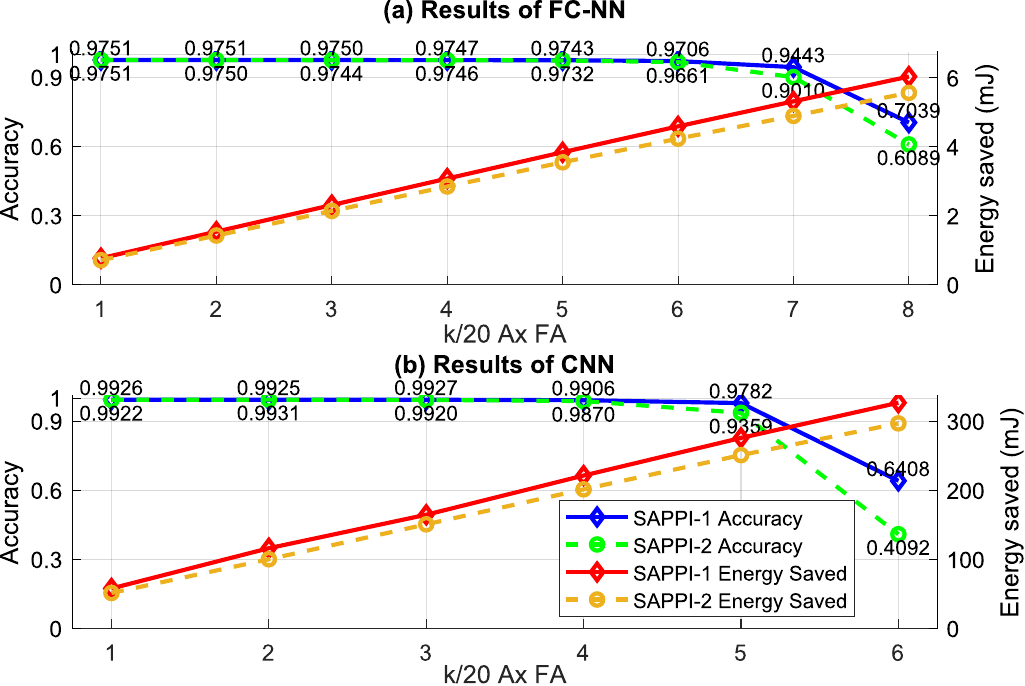}
    \caption{Results of MNIST using the proposed approximations}
    \label{fig:graph}
\end{figure}

The demand for efficient computation in machine learning (ML) applications, especially in resource-constrained environments, necessitates innovative approaches to optimize power and performance. 
In this case, we want to highlight the potential of our IMPLY-based approximations for ML applications, which was disregarded in similar approximation approaches~\cite{Seiler2023SSAxA, Fatemieh2023AFAIP, Asgari2024SAFAN}.
We verify the applicability to Fully Connected Neural Networks (FC-NN) and Convolutional Neural Networks (CNN) that are pretrained on the MNIST training dataset split (60,000 images)~\cite{Deng2012MNIST}. We utilized a partially approximated 20-bit RCA which we embedded in a shift-and-add multiplier to evaluate the networks in a post-training fashion. We used the 10,000 test images of MNIST to calculate the accuracy.
The resulting accuracy and possible energy savings for the different networks with varying approximation degrees are illustrated in \Cref{fig:graph}.

\subsection{Fully Connected Neural Network (FC-NN)}

We implemented a simple fully connected network with 28$\times$28 inputs that are connected to one hidden layer with 128 nodes and an output stage that represents the 10 possible classes. After each layer, we used a ReLU as an activation function and a normalization. Only after the last layer we used an argmax to calculate the outputs.
Our results indicate that with up to six approximated adders the accuracy for both SAPPI-1 and SAPPI-2 are maintained. SAPPI-1 has better accuracy results for all approximation degrees. With seven or more adders approximated, the accuracy rapidly degrades, rendering the network unusable. Using our approach in a fully connected neural network (FC-NN) can result in energy savings of up to 5.3 mJ (29\%) and a reduction of 23 million steps (29\%) all while preserving accuracy.

\subsection{Convolution Neural Network (CNN)}

To also show the applicability of our approaches to more complex networks, we used a CNN to classify the same problem defined before. We used a LeNet-5 inspired architecture~\cite{Lecun1998GradientLearning} for this problem since it is a classical approach to solving MNIST image classification task. It is a compact approach, compared to more modern architectures, making it the ideal example for our approximations as the impact of errors has more impact, validating our approach for more complex and error-resilient networks as well. It consists of five convolutional layers with ReLU activation functions with increasing width and height and decreasing channel dimensions.
The output of the last convolutional layer is flattened and passed through three fully connected layers with decreasing dimensions. For the exact architecture, we refer the reader to~\cite{shakibhamedan2024an,10457067}.
Our experiments indicate that with up to four adders approximated, there is no degradation in accuracy for both SAPPI-1 and SAPPI-2. With five adders approximated the accuracy for SAPPI-1 only decreases by about $1.43\%$ while the accuracy with SAPPI-2 decreases to $93.59\%$. With higher approximation degrees our methods do not lead to acceptable results. With our SAPPI-1 approach, the energy can be reduced by up to 296 mJ (21\%) while also requiring 1.33 billion fewer steps (20\%) and maintaining accuracy.

\section{Conclusion}

In this work, we presented two IMPLY-based approximate full adders in the serial topology which can drastically improve the efficiency of image processing and machine learning applications. The primary emphasis was to reduce the number of steps and energy consumption to a minimum while still being applicable in the aforementioned tasks. When embedded in partially approximated RCA, our approaches are $39\%-42\%$ more energy efficient and require $39\%-41\%$ less steps compared to exact adders. With 4/8 adders approximated, our algorithms use $10\%-13\%$ less energy and need $7\%-10\%$ fewer steps than the already efficient SoA approximations. When applied to image processing tasks our approaches can save up to 581 mJ of energy and 2.6 billion steps (cycles) in Gaussian smoothing compared to the exact adders. We demonstrated the applicability of our approximated adders in machine learning where 296 mJ and 1.3 billion steps can be saved for one inference in the MNIST dataset image classification while maintaining accuracy.

\bibliographystyle{unsrt2authabbrvpp}
\bibliography{references,memristor_lit, nima}

\begin{thebibliography}{10}

\bibitem{Boroumand2018GoogleEnergy}
A.~Boroumand \textit{et~al.}
\newblock Google workloads for consumer devices: Mitigating data movement bottlenecks.
\newblock {\em SIGPLAN Notices}, 53(2):316--331, 2018.

\bibitem{IMCTaheriNejad2024}
N.~TaheriNejad.
\newblock In-memory computing: Global energy consumption, carbon footprint, technology, and products status quo.
\newblock In {\em IEEE Nano Conference}, pp. 1--6, 2024.

\bibitem{TaheriNejad2015}
N.~Taherinejad \textit{et~al.}
\newblock Memristors' potential for multi-bit storage and pattern learning.
\newblock In {\em 2015 IEEE European Modelling Symposium (EMS)}, pp. 450--455, Oct 2015.

\bibitem{TaheriNejad2016}
N.~Taherinejad \textit{et~al.}
\newblock Fully digital write-in scheme for multi-bit memristive storage.
\newblock In {\em 2016 13th International Conference on Electrical Engineering, Computing Science and Automatic Control (CCE)}, pp. 1--6, Sept 2016.

\bibitem{Radakovits2019}
D.~Radakovits and N.~TaheriNejad.
\newblock Implementation and characterization of a memristive memory system.
\newblock In {\em 2019 IEEE 32nd Canadian Conference on Electrical and Computer Engineering (CCECE)}, pp. 1--5, May 2019.

\bibitem{Borghetti2010MemSw}
J.~Borghetti \textit{et~al.}
\newblock Memristive switches enable stateful logic operations via material implication.
\newblock {\em Nature}, 464:873--6, 04 2010.

\bibitem{Lehtonen2009StatefulIL}
E.~Lehtonen and M.~Laiho.
\newblock Stateful implication logic with memristors.
\newblock {\em 2009 IEEE/ACM International Symposium on Nanoscale Architectures}, pp. 33--36, 2009.

\bibitem{Strukov2008TheMM}
D.~B. Strukov \textit{et~al.}
\newblock The missing memristor found.
\newblock {\em Nature}, 453:80--83, 2008.

\bibitem{Gupta2018Felix}
S.~Gupta \textit{et~al.}
\newblock Felix: Fast and energy-efficient logic in memory.
\newblock In {\em 2018 IEEE/ACM International Conference on Computer-Aided Design (ICCAD)}, pp. 1--7, 2018.

\bibitem{TaheriNejad2021SIXOR}
N.~TaheriNejad.
\newblock Sixor: Single-cycle in-memristor xor.
\newblock {\em IEEE Transactions on Very Large Scale Integration (VLSI) Systems}, 29(5):925--935, 2021.

\bibitem{Huang2016tmsl}
P.~Huang \textit{et~al.}
\newblock Reconfigurable nonvolatile logic operations in resistance switching crossbar array for large-scale circuits.
\newblock {\em Advanced Materials}, 28(44):9758--9764, 2016.

\bibitem{Kvatinsky2014MAGIC}
S.~Kvatinsky \textit{et~al.}
\newblock Magic—memristor-aided logic.
\newblock {\em IEEE Transactions on Circuits and Systems II: Express Briefs}, 61(11):895--899, 2014.

\bibitem{Kvatinsky2011IMPLY}
S.~Kvatinsky \textit{et~al.}
\newblock Memristor-based imply logic design procedure.
\newblock In {\em 2011 IEEE 29th International Conference on Computer Design (ICCD)}, pp. 142--147, 2011.

\bibitem{Radakovits2021BELIEVER}
D.~Radakovits and N.~Taherinejad.
\newblock Behavioral leakage and inter-cycle variability emulator model for rerams {(BELIEVER)}.
\newblock {\em CoRR}, abs/2103.04179, 2021.

\bibitem{Liu2020AppComp}
W.~Liu \textit{et~al.}
\newblock A retrospective and prospective view of approximate computing.
\newblock {\em Proceedings of the IEEE}, 108:394--399, 03 2020.

\bibitem{Gupta2013DSP}
V.~Gupta \textit{et~al.}
\newblock Low-power digital signal processing using approximate adders.
\newblock {\em IEEE Transactions on Computer-Aided Design of Integrated Circuits and Systems}, 32(1):124--137, 2013.

\bibitem{Jiang2017AxReview}
H.~Jiang \textit{et~al.}
\newblock A review, classification, and comparative evaluation of approximate arithmetic circuits.
\newblock {\em ACM Journal on Emerging Technologies in Computing Systems (JETC)}, 13(4), 2017.

\bibitem{Chua1971MM}
L.~Chua.
\newblock Memristor-the missing circuit element.
\newblock {\em IEEE Transactions on Circuit Theory}, 18(5):507--519, 1971.

\bibitem{Rohani2017MemFA}
S.~G. Rohani and N.~TaheriNejad.
\newblock An improved algorithm for imply logic based memristive full-adder.
\newblock In {\em 2017 IEEE 30th Canadian Conference on Electrical and Computer Engineering (CCECE)}, pp. 1--4, 2017.

\bibitem{Karimi2018FullAdder}
A.~Karimi and A.~Rezai.
\newblock Novel design for a memristor-based full adder using a new imply logic approach.
\newblock {\em Journal of Computational Electronics}, 17(3):1303--1314, 2018.

\bibitem{Radakovits2020MemristiveMultiplier}
D.~Radakovits \textit{et~al.}
\newblock A memristive multiplier using semi-serial imply-based adder.
\newblock {\em IEEE Transactions on Circuits and Systems I: Regular Papers}, 67(5):1495--1506, 2020.

\bibitem{Kvatinsky2014IMPLY}
S.~Kvatinsky \textit{et~al.}
\newblock Memristor-based material implication (imply) logic: Design principles and methodologies.
\newblock {\em IEEE Transactions on Very Large Scale Integration (VLSI) Systems}, 22(10):2054--2066, 2014.

\bibitem{Fatemieh2023AFAIP}
S.~E. Fatemieh \textit{et~al.}
\newblock Fast and compact serial imply-based approximate full adders applied in image processing.
\newblock {\em IEEE Journal on Emerging and Selected Topics in Circuits and Systems}, 13(1):175--188, 2023.

\bibitem{Seiler2023SSAxA}
F.~Seiler and N.~TaheriNejad.
\newblock An imply-based semi-serial approximate in-memristor adder.
\newblock In {\em 2023 IEEE Nordic Circuits and Systems Conference (NorCAS)}, pp. 1--7, 2023.

\bibitem{Fatemieh2022AppIMC}
S.~E. Fatemieh \textit{et~al.}
\newblock Approximate in-memory computing using memristive imply logic and its application to image processing.
\newblock In {\em 2022 IEEE International Symposium on Circuits and Systems (ISCAS)}, pp. 3115--3119, 2022.

\bibitem{TaheriNejad2019SemiSerial}
N.~TaheriNejad \textit{et~al.}
\newblock A semi-serial topology for compact and fast imply-based memristive full adders.
\newblock In {\em 2019 17th IEEE International New Circuits and Systems Conference (NEWCAS)}, pp. 1--4, 2019.

\bibitem{Rohani2020SemiParallel}
S.~Ganjeheizadeh~Rohani \textit{et~al.}
\newblock A semiparallel full-adder in imply logic.
\newblock {\em IEEE Transactions on Very Large Scale Integration (VLSI) Systems}, 28(1):297--301, 2020.

\bibitem{Jiang2020AAC}
H.~Jiang \textit{et~al.}
\newblock Approximate arithmetic circuits: A survey, characterization, and recent applications.
\newblock {\em Proceedings of the IEEE}, 108(12):2108--2135, 2020.

\bibitem{Almurib2016AppDes}
H.~A. Almurib \textit{et~al.}
\newblock Inexact designs for approximate low power addition by cell replacement.
\newblock In {\em 2016 Design, Automation \& Test in Europe Conference \& Exhibition (DATE)}, pp. 660--665, 01 2016.

\bibitem{Jiang2017ARC}
H.~Jiang \textit{et~al.}
\newblock A review, classification, and comparative evaluation of approximate arithmetic circuits.
\newblock {\em ACM Journal on Emerging Technologies in Computing Systems (JETC)}, 13:1 -- 34, 2017.

\bibitem{Mittal2016ApproxComputing}
S.~Mittal.
\newblock A survey of techniques for approximate computing.
\newblock {\em ACM Computing Surveys}, 48(4):1--33, 2016.

\bibitem{Asgari2024SAFAN}
S.~Asgari \textit{et~al.}
\newblock Energy-efficient and fast imply-based approximate full adder applying nand gates for image processing.
\newblock {\em Computers and Electrical Engineering}, 113:109053, 2024.

\bibitem{Jungwirth2018SPICE}
D.~R. M.~Jungwirth and N.~TaheriNejad.
\newblock Spice implementation of vteam model, 2018.

\bibitem{knowm}
Knowm sdc memristors, knowm.org/downloads/Knowm\_Memristors.pdf.
\newblock Last accessed Feb 2024.

\bibitem{Karimi2018MemFA}
A.~Karimi and A.~Rezai.
\newblock Novel design for a memristor-based full adder using a new imply logic approach.
\newblock {\em Journal of Computational Electronics}, 17, 09 2018.

\bibitem{Khaleqi2021ImpreciseMultipliers}
M.~Khaleqi Qaleh~Jooq \textit{et~al.}
\newblock Ultraefficient imprecise multipliers based on innovative 4:2 approximate compressors.
\newblock {\em International Journal of Circuit Theory and Applications}, 49:169--184, 2021.

\bibitem{Amirafshar2023CDM}
N.~Amirafshar \textit{et~al.}
\newblock Carry disregard approximate multipliers.
\newblock {\em IEEE Transactions on Circuits and Systems I: Regular Papers}, 70(12):4840--4853, 2023.

\bibitem{Deng2012MNIST}
L.~Deng.
\newblock The mnist database of handwritten digit images for machine learning research.
\newblock {\em IEEE Signal Processing Magazine}, 29(6):141--142, 2012.

\bibitem{Lecun1998GradientLearning}
Y.~Lecun \textit{et~al.}
\newblock Gradient-based learning applied to document recognition.
\newblock {\em Proceedings of the IEEE}, 86(11):2278--2324, 1998.

\bibitem{shakibhamedan2024an}
S.~Shakibhamedan \textit{et~al.}
\newblock An analytical approach to enhancing {DNN} efficiency and accuracy using approximate multiplication.
\newblock In {\em 2nd Workshop on Advancing Neural Network Training: Computational Efficiency, Scalability, and Resource Optimization (WANT@ICML 2024)}, 2024.

\bibitem{10457067}
S.~Shakibhamedan \textit{et~al.}
\newblock Ace-cnn: Approximate carry disregard multipliers for energy-efficient cnn-based image classification.
\newblock {\em IEEE Transactions on Circuits and Systems I: Regular Papers}, 71(5):2280--2293, 2024.

\end{thebibliography}

\end{document}